\documentstyle[11pt,paspconf,epsf]{article}

\begin{document}

\title{X-ray Clusters and the Search for Cosmic Flows}

\author{P. D. Lynam, C. A. Collins, P. A. James}
\affil{Astrophysics Research Institute, Liverpool JMU, Twelve Quays House, 
Egerton Wharf, Birkenhead, CH41 1LD, United Kingdom}

\author{H. B\"{o}hringer}
\affil{Max-Planck-Institut f\"{u}r Extraterrestrische Physik, D-85740
Garching, Germany} 

\author{D. M. Neumann}
\affil{CEA Saclay, Services d'Astrophysique, Orme des Merisiers, 
B\^{a}t. 709,91191, Gif-sur-Yvette Cedex, France} 

\begin{abstract}
We are engaged in an investigation of the relationship between the properties 
of BCG candidates and X-ray characteristics of their host clusters for a 
flux-limited ($F_{X} = 3 \times 10^{-12}$ erg s$^{-1}$ cm$^{-2}$) sample of 
$\sim 250$ ACO clusters from the {\it ROSAT} all-sky survey. We aim to search 
for the convergence scale of bulk streaming flows within the $\sim 300\,h^{-1}$
Mpc defined by this sample. X-ray selection provides significant advantages 
over previous optically selected samples. No $R$-band magnitude-structure 
correlation is present in this sample. Furthermore, no correlation between 
$R$-band magnitude of the BCG candidate and X-ray luminosity of the host 
cluster is evident. The resultant scatter of $\sim$ 0.34 mag. is larger than 
in (corrected) optically selected samples. Hence, attempts to recover Local 
Group peculiar velocity vectors with respect to inertial frames defined by 
such samples via standard candle methods may have limited sensitivity.
\end{abstract}



\section{Introduction}
Efforts to improve the sensitivity of redshift-independent distance estimators
as peculiar velocity probes exploit correlations in the physical properties of
the objects used. One such correlation, between absolute magnitude, $M_{R}$ 
and Hoessel's (1980) `structure parameter', $\alpha$ ($\equiv\left.-0.921
\left[\delta M_{R}/\delta \ln r\right]\right|_{r_{m}}$; logarithmic slope of 
the luminosity profile at a metric aperture, $r_{m}$) was employed by Lauer 
\& Postman (1994, hereafter LP) for a sample of optically selected Brightest 
Cluster Galaxies (BCGs) in 119 ACO (Abell, Corwin \& Olowin 1989) hosts, to 
reduce the scatter in BCG $M_{R}$ from 0.33 to 0.24 mag. Subsequent analysis 
suggested a large-scale coherent bulk flow --- a result in conflict with 
current cosmological models (e.g. Strauss \& Willick 1995). Hudson \& Ebeling 
(1997) speculated that this scatter may be further reduced via corrections for
BCG environment. We explore this possibility with an independent, X-ray 
selected data set. 

\section{X-ray Selected Sample} 

\begin{figure}[h]
\vspace*{7.0cm}
\includegraphics{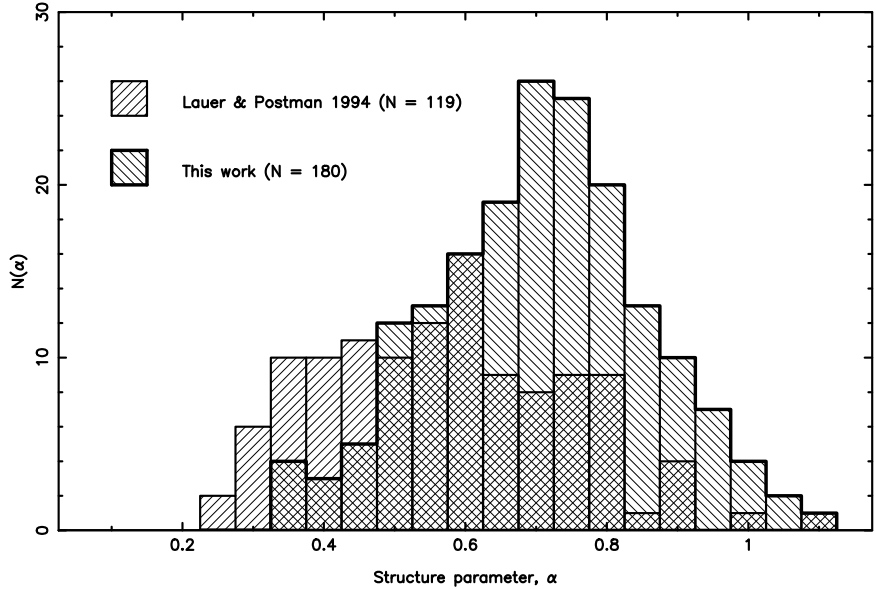}
\vspace*{0.5cm}
\caption{Structure parameter distributions for optically (LP) and X-ray 
selected (this work) BCG candidate samples.}
\label{fig2.comphist}
\end{figure}

X-ray selection has significant advantages over optical selection: ($i$) 
Diffuse X-ray emission from cluster cores identifies bona-fide clusters and 
reduces superposition effects, which in optical catalogues, can overestimate 
cluster richness. ($ii$) X-ray parameters provide a more physical reflection 
of the nature of cluster environments. ($iii$) X-ray surveys are background 
limited and thus free from problems in estimating local background galaxy 
number density. Additionally, surveying the {\it whole sky} using a single 
detector avoids biases present in optical catalogues, assembled from disparate 
survey characteristics (e.g. ACO).

We search for diffuse emission above an X-ray flux limit of $3 \times 
10^{-12}$ erg s$^{-1}$ cm$^{-2}$ in the latest reduction of the {\it ROSAT} 
all-sky survey (RASS II). The resulting database is paired with coordinates 
for ACO clusters with published redshifts, 
$z_{LG} \mathrel{\hbox{\rlap{\hbox{\lower4pt\hbox{$\sim$}}}\hbox{$<$}}} 0.1$. 
The $\sim250$ clusters for which the ACO coordinate lies within $\sim15$ 
arc-minutes of the X-ray centroid form a parent sample. Composite images 
frequently show coincidences between peaks of X-ray emission and projected 
position of dominant galaxies (Lazzati \& Chincarini 1998). BCG candidates 
for our sample are identified via this positional coincidence. In over 90\% 
of cases, positional coincidences are unambiguous. When no candidate is found 
close to a local X-ray centroid, the dominant elliptical closest to the global
cluster X-ray centroid is adopted as the BCG candidate.

For $\sim 200$ clusters, for which photometric zero point accuracy 
$\Delta R \mathrel{\hbox{\rlap{\hbox{\lower4pt\hbox{$\sim$}}}\hbox{$<$}}} 
0.03$ mag., we measure structure parameter, $\alpha$ at 10 $h^{-1}$ kpc and 
$R$-band absolute metric aperture magnitude, $M_{R}$ within elliptical 
apertures of this semi-major axis (excluding contaminating sources) of all 
ellipticals within X-ray peaks.

\section{Results: BCG Structure and Environment}

\begin{figure}[h]
\vspace*{7cm}
\includegraphics{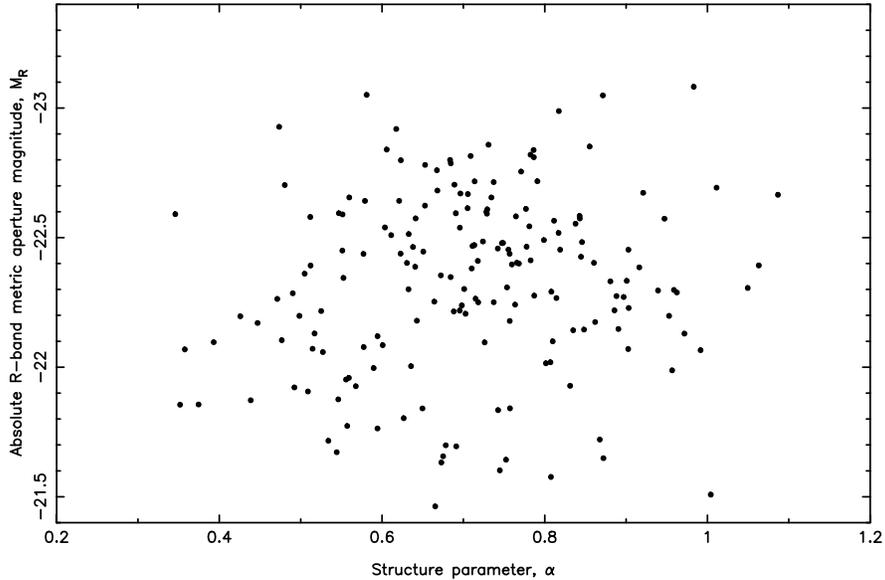}
\vspace*{0.5cm}
\caption{Absolute metric magnitude as function of structure parameter for the 
X-ray selected BCG candidate sample.}
\label{fig3.magalpha}
\end{figure}

Figure \ref{fig2.comphist} compares $\alpha$ distributions for the LP sample 
and a subset of the X-ray sample (for which positional coincidences are 
unambiguous). (a) The mean value of the X-ray distribution 
($\bar{\alpha} = 0.71$) is higher than the LP sample ($\bar{\alpha} = 0.57$). 
(b) The LP data show excess low-$\alpha$ ($ < 0.5$) galaxies. Conversely, the 
X-ray selected sample shows an excess contribution from high-$\alpha$ 
($ > 0.6$) galaxies. The distribution of X-ray selected BCG candidates appears 
shifted to high-$\alpha$ values. (c) Despite larger sample size, the X-ray 
selected distribution occupies a narrower range of $\alpha$ than the LP case. 
(d) The X-ray sample histogram is more Gaussian than the LP sample.

\begin{figure}[h]
\vspace*{8cm}
\includegraphics{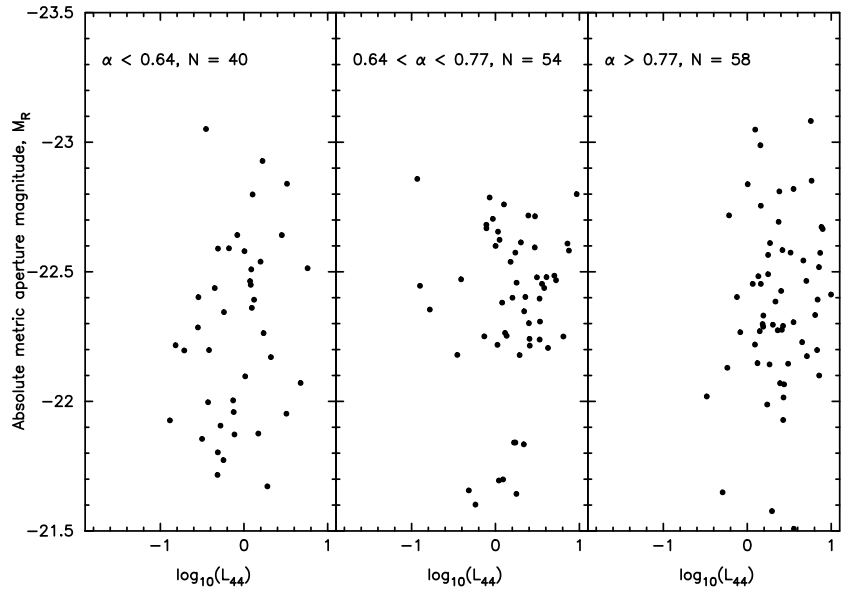}
\vspace*{0.5cm}
\caption{$M_{R}$ and $\alpha$ as function of cluster X-ray luminosity in units
of $10^{44}$ ergs s$^{-1}$, $L_{44}$ (c.f. Hudson \& Ebeling 1997 fig. 2). 
Spearman correlation coefficients for left, centre and right panels are -0.23,
-0.05 and -0.01 respectively --- all lie well outside 95\% confidence 
intervals.}
\label{fig4.triframe}
\end{figure}
  
Comparing identifications in 49 clusters common to both samples, we find 
$\sim 30\%$ of these hosts have different BCG candidates, depending upon 
selection method. In every common cluster in which LP find a low-$\alpha$ 
($ < 0.5 $) BCG, our objective technique selects a different candidate, always
with higher $\alpha$ than the optically selected BCG. Thus, X-ray selection 
reduces contaminating effects evident in optical samples, preferentially 
selecting high-$\alpha$ BCG candidates and producing a more homogeneous sample.

Figure \ref{fig3.magalpha} does not show any $M_{R}$-$\alpha$ correlation 
(c.f. LP fig. 4). Appealing to a speculated correction for environment 
(Hudson \& Ebeling 1997), we consider 152 clusters with reliably determined 
X-ray luminosities, $L_{44}$. The response of $M_{R}$ to $L_{44}$ 
(figure \ref{fig4.triframe}) does not indicate a correlation in any interval 
of $\alpha$. Hence, despite the homogeneity of X-ray selected samples, these 
candidates remain relatively poor standard candles. The scatter remains 
$\sim 0.34$ mag. dominating any dipole signal that may be present as the 
result of a coherent bulk streaming motion.

\section{Discussion}

The $M_{R}$-$\alpha$ correlation in optical BCG samples (Hoessel 1980, LP)
is principally constrained at low-$\alpha$. Above $\alpha \simeq 0.5$, 
scatter about the mean relation increases and no correlation is evident. 
X-ray selection preferentially samples this high-$\alpha$ regime. Therefore 
$M_{R}$-$\alpha$ correlations result from galaxies not coincident with X-ray 
centroids. Such galaxies are less likely to reflect the dynamics of the 
underlying cluster potential than X-ray centroid coincident BCG(s) in the same 
host. The existence of $M_{R}$-$\alpha$ correlations may signal biases in 
selection procedure via inclusion of poor tracers of cluster peculiar velocity.
Since homogeneous samples of X-ray coincident sources fail to improve BCG
reliability as standard candles, perhaps their usefulness in cosmic flow 
studies has been overstated.

\end{document}